\documentclass[aps,prl,twocolumn,noshowpacs,superscriptaddress]{revtex4-2}
\usepackage{graphicx,epsfig, subfigure}
\usepackage{amsbsy}
\usepackage{amssymb}
\usepackage{amsmath}
\usepackage{hyperref}
\usepackage[bottom]{footmisc}
\usepackage{xcolor}
\usepackage{color}
\usepackage{times}
\usepackage{textcomp}
\usepackage{verbatim} 

\begin{document}

\newcommand{\eqnref}[1]{Eq.~\ref{#1}}
\newcommand{\figref}[2][]{Fig.~\ref{#2}#1}
\newcommand{\RN}[1]{%
  \textup{\uppercase\expandafter{\romannumeral#1}}%
}

\title{Non-reciprocal energy transfer through the Casimir effect}

\author{Zhujing Xu}
	\affiliation{Department of Physics and Astronomy, Purdue University, West Lafayette, Indiana 47907, USA}
\author{Xingyu Gao}
\affiliation{Department of Physics and Astronomy, Purdue University, West Lafayette, Indiana 47907, USA}
\author{Jaehoon Bang}
\affiliation{School of Electrical and Computer Engineering, Purdue University, West Lafayette, Indiana 47907, USA}
\author{Zubin Jacob}
\affiliation{School of Electrical and Computer Engineering, Purdue University, West Lafayette, Indiana 47907, USA}
	\affiliation{Birck Nanotechnology Center,  Purdue University, West Lafayette, Indiana 47907, USA}
\author{Tongcang Li}
	\email{tcli@purdue.edu}
	\affiliation{Department of Physics and Astronomy, Purdue University, West Lafayette, Indiana 47907, USA}
	\affiliation{School of Electrical and Computer Engineering, Purdue University, West Lafayette, Indiana 47907, USA}
	\affiliation{Birck Nanotechnology Center, Purdue University, West Lafayette, Indiana 47907, USA}	
	\affiliation{Purdue Quantum Science and Engineering Institute, Purdue University, West Lafayette, Indiana 47907, USA}

\date{\today}

\begin{abstract}
{\normalsize A fundamental prediction of quantum mechanics is that there are random fluctuations everywhere in a vacuum because of the zero-point energy. Remarkably, quantum electromagnetic fluctuations can induce a measurable force between neutral objects, known as the Casimir effect \cite{Casimir1948}, which has attracted broad interests \cite{Lamoreaux1997,PhysRevLett.81.4549,Munday2009,Manjavacas2010PRL,Wilson2011,Tang2017,RevModPhys.88.045003,somers2018measurement}. The Casimir effect can dominate the interaction between microstructures at small separations and has been utilized to realize nonlinear oscillation \cite{Chan2001nonlinear}, quantum trapping \cite{Zhao984}, phonon transfer \cite{Fong2019}, and dissipation dilution \cite{Pate2020CasimirSpring}.   However, a non-reciprocal device based on quantum vacuum fluctuations remains an unexplored frontier. Here we report quantum vacuum mediated non-reciprocal  energy transfer between two micromechanical oscillators. We modulate the Casimir interaction parametrically to realize strong coupling between two oscillators with different resonant frequencies.   We engineer the system's spectrum to have an exceptional point \cite{Mirieaar7709ScienceEP,Berry2004,heiss2012physics,Bender1998PT} in the parameter space and observe the asymmetric topological structure near it. 
By dynamically changing the parameters near the exceptional point and utilizing the non-adiabaticity of the process, we achieve non-reciprocal energy transfer with high contrast. 
Our work represents an important development in utilizing quantum vacuum fluctuations to regulate energy transfer at the nanoscale and build functional Casimir devices.
}
\end{abstract}

\maketitle


In 1948,  Casimir speculated that two uncharged metal plates separated by a
vacuum gap would experience an attractive force due to
quantum vacuum fluctuations \cite{Casimir1948}.
Besides its fascinating origin  and importance in  fundamental physics   
 \cite{mostepanenko1997casimir}, the Casimir effect can  dominate at sub-micrometer distances and is essential for micro and nano technologies
 \cite{rodriguez2011casimir,Zhao2003}. Several studies have demonstrated Casimir effect-based devices with various functions such as nonlinear oscillation \cite{Chan2001nonlinear}, quantum trapping \cite{Zhao984}, phonon transfer \cite{Fong2019,PendryPRB2016}, and dissipation dilution \cite{Pate2020CasimirSpring}. We note that many essential devices such as diodes, isolators, and circulators require non-reciprocity. However, a non-reciprocal device based on the Casimir effect  has been elusive.  

\begin{figure}[t]
	\centerline{\includegraphics[width=1.0\linewidth]{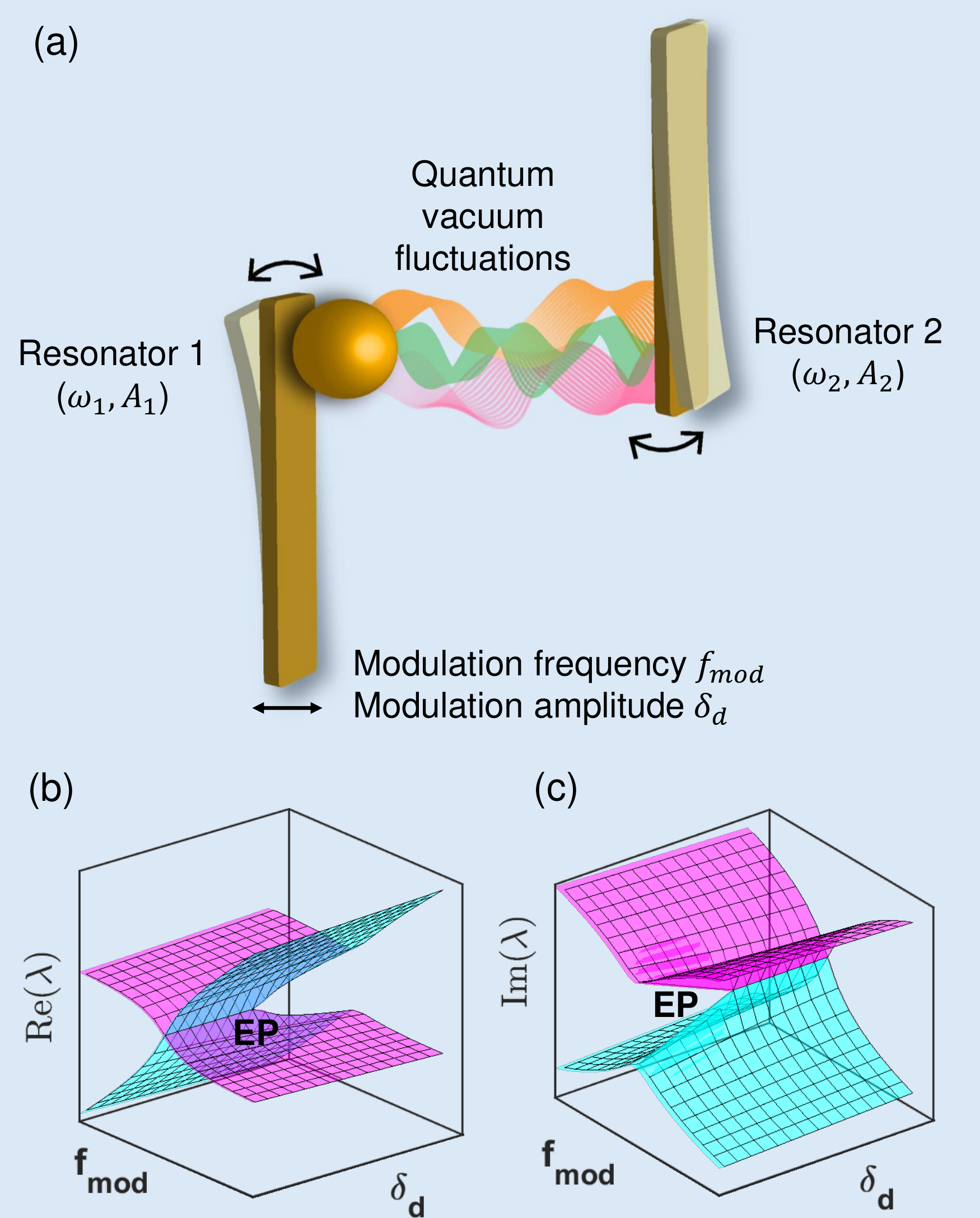}}
	\caption{\textbf{Casimir effect in the dual-cantilever system and  eigenvalues near the exceptional point.} (a): Two modified cantilevers with resonant frequencies $\omega_1$ and $\omega_2$ experience a Casimir force due to quantum vacuum fluctuations. The vibration amplitudes of two cantilevers are denoted as $A_1$ and $A_2$. An additional slow modulation with a frequency $f_{mod}$ and an amplitude $\delta_d$ is applied on resonator 1 to realize parametric coupling.  (b) and (c): The real part (Re$(\lambda)$) and the imaginary part (Im$(\lambda)$) of the eigenvalues $\lambda_{\pm}$ of the system Hamiltonian  are shown as a function of the modulation frequency $f_{mod}$ and the modulation amplitude $\delta_d$. The two eigenvalues   exhibit a nontrivial topological structure near the exceptional point (EP). }
	\label{setup}
\end{figure}

 Similar to the control of electric current with diodes, we develop an efficient ``Casimir diode'' that can rectify energy transfer coupled by Casimir interaction. The non-reciprocity is realized by dynamic modulation of the nonlinear Casimir interaction near an exceptional point \cite{Mirieaar7709ScienceEP,Berry2004,heiss2012physics,Bender1998PT}.
An exceptional point is a branch singularity of a non-Hermitian system such that the eigenvalues of the system Hamiltonian collapse with each other for both real and imaginary parts in the parameter space \cite{Berry2004,Berry_2011}.  The exceptional point has attracted broad interests in optics, optomechanics, and acoustics \cite{Regensburger2012,zhao2018exceptional,Xu2016,Doppler2016,Chen2017,hadad2020possibility} since it exhibits a unique topological structure. In particular, non-reciprocal energy transfer between two optically coupled mechanical modes of a membrane inside an optical cavity has been carried out by dynamically controlling the parameters in a loop that encloses an exceptional point \cite{Xu2016}. Such non-reciprocal operations open new directions for controlling optomechanical systems. However, it has not been realized with Casimir interactions before. We utilize the strong nonlinearity of the Casimir interaction and asymmetric structure near the exceptional point to break the time reversal symmetry by modulating the separation between two micromechanical resonators at the desired frequency and amplitude (Fig. \ref{setup}). In this way, we realize non-reciprocal energy transfer with the Casimir interaction. The direction of energy transfer depends on the sequence of operations. Therefore, the system provides the flexibility for future applications in Casimir-based devices.

\begin{figure*}[t]
	\centerline{\includegraphics[width=1.0\linewidth]{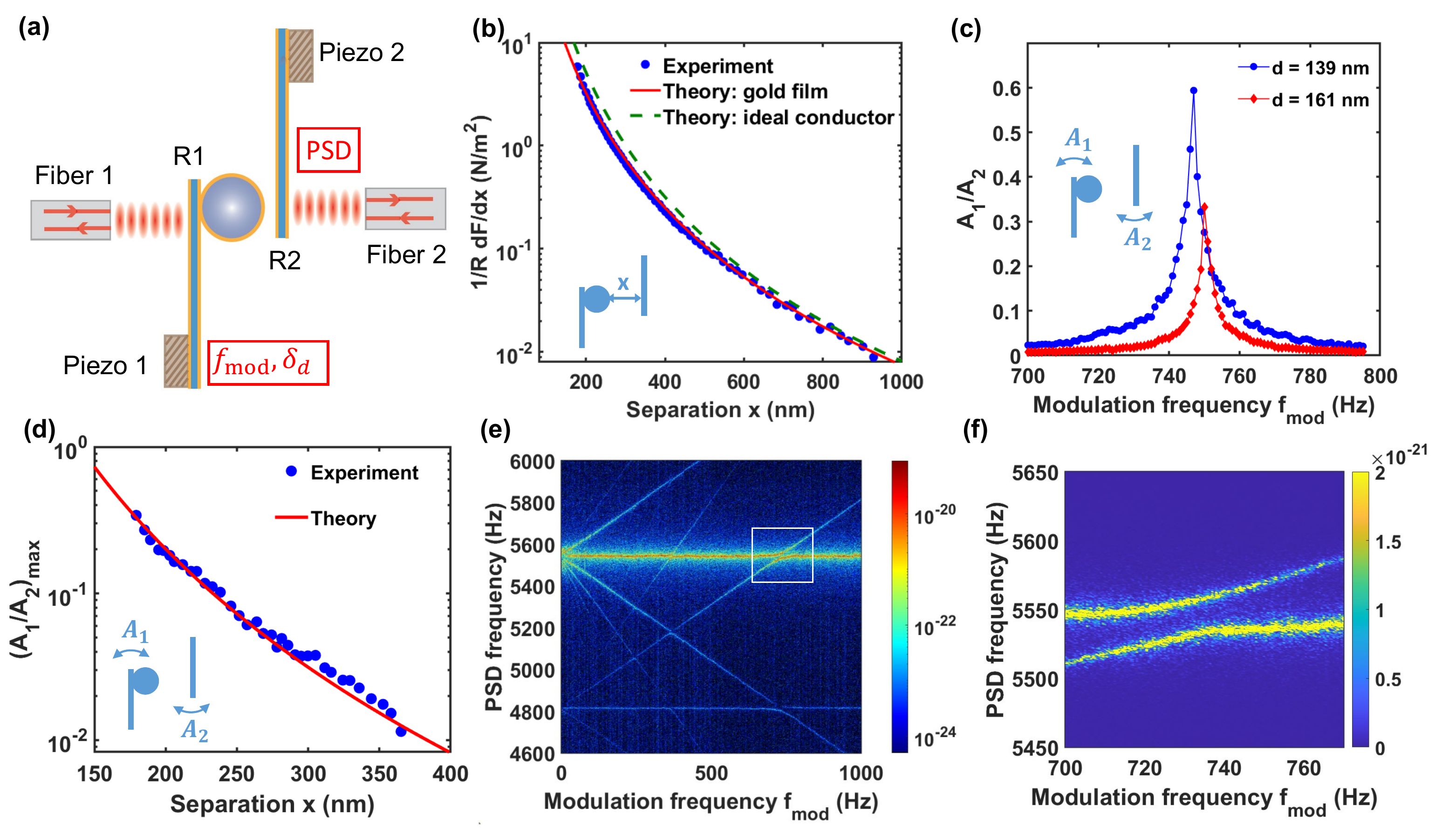}}
	\caption{\textbf{Force measurement and energy transfer by the Casimir effect.} (a). The schematic of the dual-cantilever fiber interferometer setup. A  microsphere is attached to one cantilever and both cantilevers are coated with gold  for good conductivity and reflectivity. Their motions are monitored by two fiber interferometers independently. Two piezo chips at the end of cantilevers are used to drive the cantilevers and change the separation between two cantilevers. (b). The measured force gradient divided by the radius of the microsphere is shown as a function of the separation. The red solid line is the theoretical prediction for real gold material. The green dashed curve is the theoretical prediction for ideal conductor. (c).The ratio of the displacement of two cantilevers is shown as a function of the modulation frequency $f_{mod}$ at a separation of 139 nm and 161 nm. (d). Parametric modulation is applied to couple two cantilevers. The ratio of the displacement of two cantilevers is shown as a function of the separation when cantilever 2 is first excited and the modulation frequency equals to the frequency difference between two cantilevers. The red solid curve is the theoretical prediction. (e). Power spectral density (PSD) of the cantilever 2 as a function of the modulation frequency $f_{mod}$ and PSD frequency. (f) is the refined scan of the white box shown in (e).}
	\label{energy_transfer}
\end{figure*}

Our device consists of two cantilevers with resonant frequencies $\omega_1$ and $\omega_2$ as shown in  Fig.\ref{setup}(a). A microsphere is attached to the left cantilever. The oscillating amplitudes of two cantilevers are denoted as $A_1$ and $A_2$. Two cantilevers experience quantum vacuum fluctuations and attract each other by the Casimir force. We apply an additional modulation of the separation  to couple two cantilevers with different frequencies. For the separation considered in our experiment, quantum fluctuations (instead of thermal fluctuations) dominate the Casimir interaction (see Supplementary Fig. S2). 
The schematic of the experimental setup is shown in Fig. \ref{energy_transfer}(a). The motion of two cantilevers are monitored independently by two fiber interferometers.  The natural frequency and damping of two modified cantilevers are $\omega_1 = 2\pi\times 4826$ Hz, $\omega_2 = 2\pi\times 5582$ Hz, $\gamma_1 = 2\pi\times 2.65$ Hz, and $\gamma_2 = 2\pi\times 2.68$ Hz when they are far away. The microsphere and both cantilevers are coated by   gold. 
When two surfaces are close to each other, they will experience the Casimir interaction. 
At small separations,  the Casimir force  between an ideal conductive sphere and an ideal conductive plate is \cite{Chan2001}
\begin{equation}
F_C^0(x)=-\frac{\pi^3 \hbar c}{360}\frac{R}{x^3},
\label{eq_Casimir}
\end{equation}
where $R$ is the radius of the sphere, $x$ is the separation between the sphere and the plate, $\hbar$ and $c$ are the reduced Planck constant and the speed of light, respectively. The Casimir force $F_C$ between real materials can be calculated by the Lifshitz theory \cite{Lifshitz:1956} and more details can be found in the Supplementary Information.
The measured Casimir force gradient divided by the radius of the microsphere is shown in Fig.\ref{energy_transfer}(b). The experimental data agrees well with the theoretical prediction for the Casimir force between real gold films under the proximity-force-approximation.

We now discuss how we achieve strong coupling between two cantilevers by quantum vacuum fluctuations.
To couple  two cantilevers with different resonant frequencies, we modulate the separation between them  at a slow rate $\omega_{mod}$. 
The effective coupling strength is controllable by changing the  modulation amplitude $\delta_d$. Different from direct coupling that requires identical resonant frequencies, parametric coupling gives us more freedom to couple arbitrary resonators  and control the coupling time, the coupling strength, and the effective detuning of two oscillators  \cite{Huang2013,Mathew2016}. Motion transduction and parametric coupling between two mechanical resonators has been realized by modulating the electrostatic interaction parametrically \cite{Huang2013}.  Tunable intermodal coupling has been carried out on graphene based resonators by modulating the resonator frequency parametrically \cite{Mathew2016}. 
If we resonantly excite resonator 2 with a constant amplitude $A_2$ and modulate the separation at a rate of $f_{mod}=f_{21}$, where $f_{21}= (\omega_2-\omega_1)/2\pi$, the excitation on resonator 2 is down-converted to the vibration of resonator 1 (Fig.\ref{energy_transfer}(c)) \cite{Mahboob2012}. 
Under weak-coupling approximation, the ratio of the oscillating amplitudes between two resonators ($A_1$ and $A_2$) at their own resonant frequencies at the steady state  is given as
\begin{equation}
\frac{A_1}{A_2} = |\frac{d^2F_{C}}{dx^2}| \frac{\omega_1 \delta_d}{2\gamma_1 k_1}.
\label{eq_ratio}
\end{equation} Note this coupling is due to the second derivative of the Casimir force $d^2F_{C}/dx^2$. For a spring force $F_s \propto -x$, the second derivative will be 0 and the parametric coupling can not be achieved by simply modulating the separation.
 The maximum transduction amplitude $A_1/A_2$ as a function of separation is shown in Fig.\ref{energy_transfer}(d).   The experimental result agrees with the prediction  Eq. \ref{eq_ratio}. Thus we have realized the energy transfer through quantum vacuum fluctuations. We can reduce the separation $x$ or increase the modulation amplitude $\delta_d$ to achieve strong coupling.
Fig.\ref{energy_transfer}.(e) shows the power spectral density (PSD) of cantilever 2 as a function of the PSD frequency and modulation frequency $f_{mod}$ when parametric modulation is applied on cantilever 1 (left).  Near resonance, level repulsion is observed in Fig. \ref{energy_transfer}(f), which shows strong coupling between cantilevers. More details of the spectrum can be found in Supplementary Fig. S8.

\begin{figure*}
	\centerline{\includegraphics[width=1.0\linewidth]{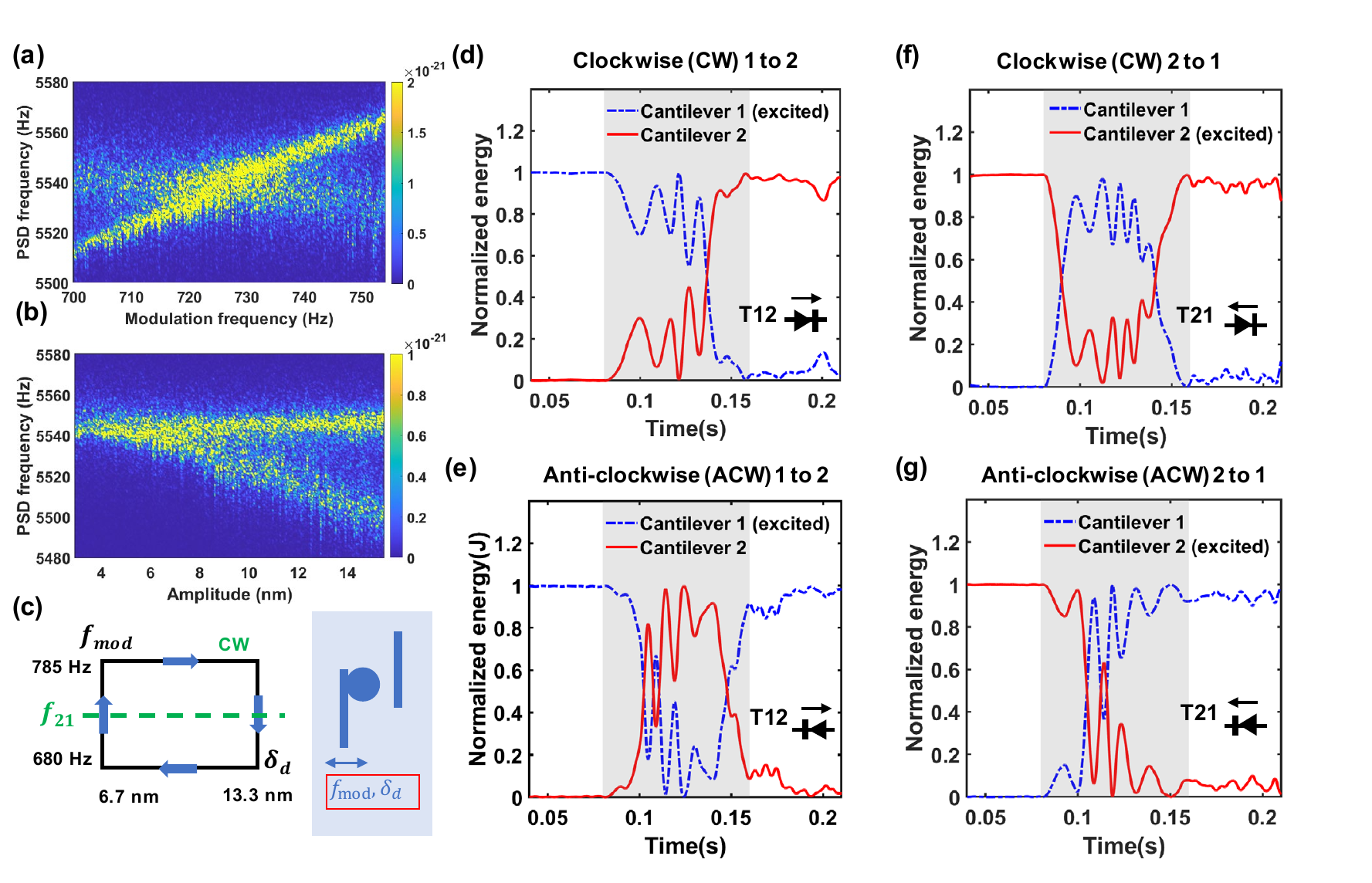}}
	\caption{\textbf{Non-reciprocal energy transfer by the Casimir effect}. (a). PSD intensity of cantilever 2 as a function of modulation frequency $f_{mod}$ and PSD frequency when $\delta_d = 5.5 \pm 0.6$ nm. (b). PSD intensity of cantilever 2 as a function of modulation amplitude $\delta_d$ and PSD frequency when $f_{mod} = 727$ Hz. (c). A clockwise (CW) control loop in the parameter space for the non-reciprocal energy transfer. (d)-(g). Measurement of the normalized energy in the transfer process. One cantilever is first driven to the excited state from 0 ms to 80 ms. The dynamical control starts at 80 ms (shaded in gray) for clockwise (CW) and anti-clockwise (ACW) loop and lasts for 80 ms. A clockwise loop allows the energy transferred from 1 to 2 and avoids the reverse direction. On the other hand, an anti-clockwise loop allows the energy transferred from 2 to 1 and avoids the reverse direction. 
	}
	\label{Topological_energy}
\end{figure*}

In the experiment, non-reciprocal energy transfer is realized by dynamically controlling the parameters in a loop near the exceptional point. Parametric coupling gives us the freedom to control coupling strength and detuning of the system as a function of time. 
In an interaction picture, the simplified Hamiltonian (see Methods and Supplementary Information) is  
\begin{eqnarray}
H_{int} = \begin{pmatrix}
-\frac{i\gamma_1}{2} & \frac{g}{2} \\ \frac{g}{2} & -\frac{i\gamma_2}{2}-\delta 
\end{pmatrix},
\label{Hamiltonian0}
\end{eqnarray}
where $\gamma_{1,2}$ is the damping rates of two cantilevers and $g$ is the coupling strength between two cantilevers. $\delta$ is the detuning of the system.  The coupling strength and detuning are directly related to the modulation amplitude $\delta_d$ and modulation frequency $f_{mod}=\omega_{mod}/2\pi$ as $g = \frac{d^2F_C}{dx^2}\delta_d/2\sqrt{m_1m_2\omega_1\omega_2}$ and $\delta = 2\pi (f_{mod}-f_{21})$. Here $\frac{d^2F_C}{dx^2}$ is the second derivative of Casimir force $F_C$ and $m_{1,2}$ is the mass of two cantilevers. For this open system, the exceptional point is located at $g = \frac{|\gamma_1-\gamma_2|}{2}$ and $\delta = 0$. At this specific point, the two  eigenvalues $\lambda_{\pm}$ of the Hamiltonian are degenerate for both real and imaginary parts as shown in Fig.\ref{setup}(b) and \ref{setup}(c). The eigenvalues (two surfaces) intersect each other and exhibit a nontrivial topological structure near the exceptional point. The asymmetric structure near the exceptional point provides us freedom to construct dynamical operations that break the time reversal symmetry.
Experimentally we can control the modulation frequency $f_{mod}$ and the modulation amplitude $\delta_d$ independently as a function of time and realize non-reciprocal energy transfer by controlling the parameters along a  loop near the exceptional point.

We first measure the spectrum of the system near the exceptional point experimentally. The exceptional point locates at the point when two eigenvalue surfaces intersect each other. 
Under the natural condition, two cantilevers have comparable damping rates and the damping difference is close to zero. Thus the  exceptional point locates at $\delta_d = 0$, which means no coupling between two cantilevers. To break symmetry and achieve non-reciprocal energy transfer, we need to add extra gain or loss to the system to shift the exceptional point. Here we add extra loss to one of the cantilevers for simplicity. In the experiment, we apply an additional damping on cantilever 2 (right) such that $\gamma_2 = 2\pi\times 13.82$ Hz. 
Fig.\ref{Topological_energy} (a) and (b) are the measured PSD of cantilever 2 and they show that the exceptional point in this double-cantilever system is located approximately at $\delta_d = 5.5$ nm and $f_{mod} = 727$ Hz. The separation is calculated to be 76~nm from the coupling strength. Based on the measurement of exceptional point, we design a dynamical clockwise control loop as shown in Fig. \ref{Topological_energy}(c).  The modulation amplitude is continuously controlled from $6.7\pm 0.6$ nm to $13.3\pm 0.6$ nm and the modulation frequency is tuned from 680 Hz to 785 Hz. 
The total loop time is 80~ms. When the modulation frequency $f_{mod}$ is tuned from 680 Hz to 785 Hz and $\delta_d = 6.7$ nm, the minimum energy gap between two eigenstates  is small compared to the operation speed and hence the process is non-adiabatic. The energy prefers to be transferred to the  cantilever with less damping after going through the control process when $\delta_d = 6.7$ nm. The process is adiabatic for other parts of the control loop since the energy gap between two eigenstates is large compared to the operation speed.  Therefore, time reversal symmetry is broken by the control loop shown in Fig.\ref{Topological_energy}(c). An important striking feature is that the preferred transfer direction of energy transfer depends on the direction of the control loop. Our simulations (see Supplementary
Information) of the energy transfer process take into account
the nonlinearity inherent in this Casimir force coupled system.

\begin{figure}
	\centerline{\includegraphics[width=1.0\linewidth]{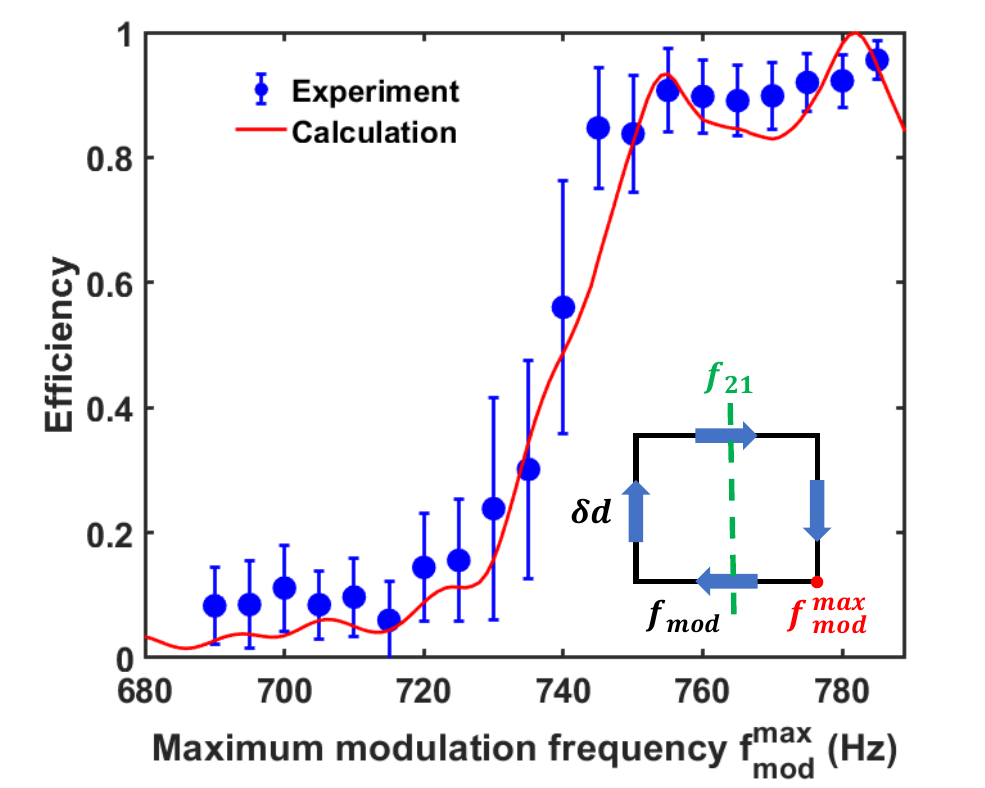}}
	\caption{\textbf{Efficiency of the non-reciprocal energy transfer by the Casimir effect.} The transfer efficiency as a function of the maximum modulation frequency $f^{max}_{mod}$ for the clockwise loop when cantilever 1 is first excited. The minimum modulation frequency is fixed at 680 Hz and the modulation amplitude is changing from $6.7\pm 0.6$ nm to $13.3\pm 0.6$ nm. When the control loop includes the resonant modulation frequency $f_{mod}=f_{21}$ (as shown in the green dashed line), energy is transferred from cantilever 1 to cantilever 2. Little energy is  transferred when the maximum modulation frequency is lower than $f_{21}$. The error bars show standard deviations of data points.} 
	\label{Efficiency}
\end{figure}

We now realize a Casimir diode by engineering the system dynamically. Figure \ref{Topological_energy}(d)-(g) show the experimental results of the non-reciprocal energy transfer process for the clockwise (CW) and the anti-clockwise (ACW) loops with different cantilevers being excited. We use normalized energy, which is defined as $E_1/(E_1+E_2)$ and $E_2/(E_1+E_2)$, to quantify the portion of energy for two cantilevers in the transfer process. Here $E_1$ and $E_2$ are the energy of two cantilevers by measuring their oscillating amplitudes. 
One of the cantilevers is first driven resonantly to the excited state from 0 to 80 ms. The dynamical control loop is then applied starting at 80 ms as shown in the gray shaded area. 
We measure the energy $E_1$ and $E_2$ at the end of the loop ($t=160$ ms) and see whether energy is transferred to a different state. 
Fig. \ref{Topological_energy}.(d) and (f) show that a clockwise loop allows energy transferring from 1 to 2 while avoids the opposite direction. Cantilever 2 dominates the total energy in the system after the clockwise loop no matter what the initial state is. 
On the contrary, cantilever 1 dominates the energy after the anti-clockwise loop as shown in Fig. \ref{Topological_energy}(e) and (g). Two energy transfer processes with different directions for the same control loop have a high contrast. The simulated transfer process is shown in Supplementary Fig. S9, which shows the same key features.
 Our system is flexible to manipulate the preferable transfer direction by designing the control loop.

We demonstrate that the asymmetric structure near the exceptional point leads to highly efficient directionality. To quantify the efficiency of energy transfer after the dynamical control loop, we calculate the transfer efficiency which is defined as $\eta = E_2/(E_1+E_2)$ at the end of the control loop for the case that cantilever 1 is first excited. In this way, we can measure how much energy is transferred from cantilever 1 to cantilever 2. Similar definition applies to the reverse direction.
Fig.\ref{Efficiency} shows the dependence of transfer efficiency on the size of the control loop when cantilever 1 is first excited.
We change the maximum modulation frequency $f^{max}_{mod}$ while keeping the minimum modulation frequency at 680 Hz for a clockwise control loop as shown in the inset of Fig.\ref{Efficiency}. 
The control on the modulation amplitude $\delta_d$ remains the same as mentioned above.  When the control loop includes the resonant modulation frequency $f_{mod}=f_{21}$, energy is transferred to cantilever 2 when we first excite cantilever 1. On the contrary, when the loop does not include $f_{21}$, little energy is transferred from cantilever 1 to cantilever 2. 

In conclusion, we report non-reciprocal energy transfer with quantum vacuum fluctuations. Strong phonon coupling and energy transfer between two mechanical resonators are realized by parametric modulation. 
Under the careful design of the control loop, non-reciprocal energy transfer with high contrast is observed. Our work develops a flexible and robust method to regulate vacuum fluctuations and build functional Casimir devices. The Casimir coupling due to virtual photons can replace the optomechanical coupling due to real photons in cavity optomechanics \cite{Xu2016} for various functions. If the mechanical oscillators are cooled to the quantum regime at low temperature, the non-reciprocal Casimir effect can be used to regulate the flow of single phonons for quantum information processing.

%

\newpage

\section*{Methods}

\textbf{Experimental set-up.}
We use two AFM (atomic force microscope) cantilevers to construct our Casimir interaction system. The left cantilever has a size of $450 \, \mu{\rm m}\times  50 \, \mu{\rm m} \times 2 \, \mu{\rm m}$. A polystyrene sphere with a diameter of $69.1\pm 0.9$ $\mu$m is attached on the free end of the left cantilever by a vacuum-compatible conductive epoxy. The right cantilever has a size of $500 \, \mu{\rm m} \times 100 \, \mu{\rm m} \times 1 \, \mu{\rm m}$. 70-nm thick gold layers are coated on both sides of cantilevers and the sphere to create metallic surfaces for the Casimir interaction and a better reflectivity for detection. Films with nearly equal thicknesses are coated on both sides of cantilevers in order to reduce strains on the cantilevers. Each cantilever is mounted on a piezoelectric stack to control its equilibrium position and vibration. One cantilever is connected to the electrical ground. A biased voltage is applied on the other cantilever to minimize the effects of the patch potential (see Supplementary Fig. S5, S6). 

The motion of two cantilevers are monitored by two fiber interferometers. A 50-$\mu$W laser with a wavelength of 1310 nm is incident on the back side of a cantilever from a fiber 
and gets reflected back. The reflected light is directed into the same optical fiber and interferes with the reflected light from the fiber-air interface. The interference signal gives the information of the separation between the cantilever and the fiber and hence we can measure the motion of the cantilever. The jacket and the cladding of the fibers have been removed and the bare fiber has a diameter of 125 $\mu$m. The fibers are placed at a distance of more than 200 $\mu$m from the cantilevers to reduce the electrostatic effect. To minimize the environmental impact, the system is placed on top of optical tables by two stages of pneumatic vibration isolation (see Supplementary Fig. S1). The experiment is conducted under a pressure of $10^{-4}$ torr at room temperature.  To measure the Casimir force, we monitor the frequency shift of the cantilever at each separation using a phase-lock loop (see Supplementary Fig. S5).

\textbf{Casimir force.} We use the Lifshitz theory to calculate the Casimir force between real materials \cite{Lifshitz:1956}. The Casimir energy per unit area of two parallel plates at zero temperature with a finite separation $x$ is \cite{Lifshitz:1956}
\begin{eqnarray}
E_0(x) &=& \frac{\hbar}{4\pi^2}\int_{0}^{\infty} k_{\perp}dk_{\perp}\int_{0}^{\infty}d\xi\{\ln [1-r_{TM}^2(i\xi,  k_{\perp})e^{-2xq}]\nonumber\\
& & +\ln[1-r_{TE}^2(i\xi,k_{\perp})e^{-2xq}]\},
\label{CasimirEnergy_0K}
\end{eqnarray}
where $\xi$ is the imaginary frequency and $k_{\perp}$ is the wave vector parallel to the surface. $r_{TE}(i\xi,k_{\perp})$ and $r_{TM}(i\xi,k_{\perp})$ are  the reflection coefficients of the transverse electric and magnetic modes \cite{Lifshitz:1956}. At a finite temperature $T$, both quantum fluctuations and thermal fluctuations contribute to the Casimir energy per unit area $E(x,T)$. In our experiment, the radius of the microsphere and the dimensions of the bare cantilever are far larger than the separation. Therefore, we can use proximity-force approximation (PFA) to evaluate the Casimir force between a sphere and a plate. The Casimir force for our system is $F_C(x,T) = -2\pi R E(x,T)$, where $R$ is the radius of the microsphere.
As shown in Supplementary Fig. S2, the contribution from thermal fluctuations at $T=300$ K   is less than 6\% when the separation is less than 1000 nm, and is only about 2\% when the separation is 200 nm. Thus the effects of quantum vacuum fluctuations dominate in our experiment. As shown in Supplementary Fig. S3 and S4, the Casimir pressure between 70-nm-thick gold films is different from the Casimir pressure between infinitely-thick gold plates by less than 0.1\%, which is negligible for our experiment. The AFM image in Supplementary Fig. S7 shows the rms roughness of the gold film is about 0.8 nm, which is  small compared to the separation in our experiment.

\textbf{Effective Hamiltonian.} Under a slow modulation and Casimir interaction, the separation between two cantilevers is time-dependent such that $x(t) = d_0+\delta_d \cos(\omega_{mod}t)+x_1 (t)-x_2 (t)$. Here $d_0$ is the equilibrium separation when there is no modulation applied, $\delta_d$ is the modulation amplitude. $\omega_{mod}= 2 \pi f_{mod}$, where $f_{mod}$  is the modulation frequency.  $x_1 (t)$ and $x_2 (t)$ describe  vibrations of the cantilevers near their equilibrium positions. The motions of the cantilevers follow equations
\begin{eqnarray}
m_1\ddot{x_1}+m_1\gamma_1\dot{x_1}+m_1\omega_1^2x_1 = F_{C}(x(t)),\hspace{0.3cm}\nonumber\\
m_2\ddot{x_2}+m_2\gamma_2\dot{x_2}+m_2\omega_2^2x_2 = -F_{C}(x(t)).
\end{eqnarray}
When the modulation amplitude and the oscillation amplitude of two cantilevers are far smaller than the separation such that $\delta_d, x_1, x_2 \ll d_0$, we can expand the Casimir force term $F_{C}(d_0+\delta_d \cos(\omega_{mod}t)+x_1-x_2)$ to the second order.
Since two cantilevers have a frequency difference over 700 Hz, the direct coupling is neglected. The zero-order and first-order terms  shift the frequency of two cantilevers but have no contribution to energy transfer since they are off-resonant. The  contribution comes from the term $\frac{d^2 F_{C}}{d x^2}|_{d_0}\delta_d\cos(\omega_{mod}t)(x_1-x_2)$. Therefore, we can rewrite the equations as
\begin{eqnarray}
\ddot{x_1}+\gamma_1\dot{x_1}+\omega_1^2x_1 = \frac{\Lambda}{m_1}\cos(\omega_{mod}t)(x_1-x_2),\nonumber\\
\ddot{x_2}+\gamma_2\dot{x_2}+\omega_2^2x_2 = \frac{\Lambda}{m_2}\cos(\omega_{mod}t)(x_2-x_1),
\label{equationofmotion2}
\end{eqnarray}
where we have $\Lambda = \frac{d^2F_C}{dx^2}\delta_d$.

Here we solve the second-order ordinary equations in Eq.(\ref{equationofmotion2}) by generalizing the displacements $x_1(t)$ and $x_2(t)$ to complex values $z_1(t)$ and $z_2(t)$ such that $x_1(t) = Re[z_1(t)]$ and $x_2(t) = Re[z_2(t)]$.
We separate the fast-rotating term and the slow-varying term for $z_1(t)$ and $z_2(t)$ such that
\begin{eqnarray}
z_1(t) = A_1(t)e^{-i\omega_1t},\nonumber\\
z_2(t) = A_2(t)e^{-i\omega_2t},
\end{eqnarray}
where $A_1(t)$ and $A_2(t)$ are slow-varying amplitudes and we can neglect their second derivative terms $\ddot{A_1}(t)$ and $\ddot{A_2}(t)$ in the equations of motion. Besides, we consider the condition that the damping rate of two cantilevers are far smaller than the resonant frequency such that $\gamma_1\ll\omega_1$ and $\gamma_2\ll\omega_2$. 
Therefore, the equations of motion can be rewritten as 
\begin{eqnarray}
-i\omega_1\gamma_1A_1(t)e^{-i\omega_1t}-2i\omega_1\dot{A_1}(t)e^{-i\omega_1t}\hspace{2.6cm} \nonumber\\= \frac{\Lambda}{2m_1}(A_1(t)e^{-i(\omega_1+\omega_{mod})t}-A_2(t)e^{-i(\omega_2-\omega_{mod})t}),\hspace{0.5cm}\nonumber\\
-i\omega_2\gamma_2A_2(t)e^{-i\omega_2t}-2i\omega_2\dot{A_2}(t)e^{-i\omega_2t} \hspace{2.6cm}\nonumber\\= \frac{\Lambda}{2m_2}(A_2(t)e^{-i(\omega_2-\omega_{mod})t}-A_1(t)e^{-i(\omega_1+\omega_{mod})t}),\hspace{0.5cm}
\end{eqnarray}
where we have taken the rotating frame approximation and neglected the fast-rotating term.
Now we apply the transformation such that $A_1'(t) = A_1(t)$ and $A_2'(t) = A_2(t)e^{i\delta t}$, where $\delta = \omega_1+\omega_{mod}-\omega_2$. Then the equation of motion becomes 
\begin{equation}
	i\begin{pmatrix}\dot{A_1'}(t)\\ \dot{A_2'}(t)\end{pmatrix} = \begin{pmatrix} -i\frac{\gamma_1}{2} & \frac{\Lambda}{4m_1\omega_1}\\ \frac{\Lambda}{4m_2\omega_2} & -i\frac{\gamma_2}{2}-\delta \end{pmatrix}\begin{pmatrix}A_1'(t) \\ A_2'(t)\end{pmatrix},
\end{equation}
where we have neglected the fast-rotating terms. The vibrations of two cantilevers can be quantized as phonons. Introducing normalized amplitudes $c_1=\sqrt{\frac{m_1 \omega_1}{\hbar}} A_1'$ and $c_2=\sqrt{\frac{m_2 \omega_2}{\hbar}} A_2'$, we obtain the equation of motion for phonon modes 
\begin{equation}
i\begin{pmatrix}\dot{c_1}\\\dot{c_2}\end{pmatrix} = \begin{pmatrix} -i\frac{\gamma_1}{2} & \frac{g}{2}\\ \frac{g}{2} & -i\frac{\gamma_2}{2}-\delta \end{pmatrix}\begin{pmatrix}c_1\\c_2\end{pmatrix},
\end{equation}
where $g = \frac{\Lambda}{2\sqrt{m_1m_2\omega_1\omega_2}} = \frac{d^2 F_C}{dx^2}\delta_d\frac{1}{2\sqrt{m_1m_2\omega_1\omega_2}}$ and $\delta = \omega_1+\omega_{mod}-\omega_2$. 
 Thus the effective Hamiltonian of the system is Eq. (\ref{Hamiltonian0}) in the main text.
The eigenvalues of the Hamiltonian are 
\begin{eqnarray}
\lambda_{\pm} = -\frac{\delta}{2}-i\frac{\gamma_1+\gamma_2}{4}\hspace{5cm}\nonumber\\
\pm\frac{1}{2}\sqrt{-\frac{(\gamma_1-\gamma_2)^2}{4}+\delta^2+g^2-i(\gamma_1-\gamma_2)\delta}.\hspace{1cm}
\end{eqnarray}
The eigenvalue depends on modulation amplitude $\delta_d$ and modulation frequency $\omega_{mod}$.
We can also see that the exceptional point locates at $\delta = 0$ and $g = \frac{|\gamma_1-\gamma_2|}{2}$ which means that $\omega_{mod} = \omega_2-\omega_1$ and $\delta_d =\frac{|\gamma_1-\gamma_2|\sqrt{m_1m_2\omega_1\omega_2}}{d^2F_C/dx^2} $. 




\section*{Acknowledgments}
We are grateful to supports from the Defense Advanced Research Projects Agency (DARPA) NLM program, and the Office of Naval Research under Grant No. N00014-18-1-2371.

\section*{Author contributions}
T.L., Z.X. and Z.J. conceived and designed the project. Z.X., T.L., X.G. and J.B. built the setup. Z.X. performed measurements. Z.X. and X.G. performed calculations. T.L. and Z.J. supervised the project. All authors contributed in data analysis and writing of the manuscript.


\end{document}